\documentclass[10pt,a4paper]{article}

\ifdefined\pdfinfoomitdate\pdfinfoomitdate=1\fi
\ifdefined\pdftrailerid\pdftrailerid{}\fi
\ifdefined\pdfsuppressptexinfo\pdfsuppressptexinfo=-1\fi

\usepackage[utf8]{inputenc}
\usepackage[T1]{fontenc}
\usepackage[english]{babel}
\usepackage{geometry}
\usepackage[scaled=0.95]{helvet}

\usepackage{microtype}
\usepackage{setspace}
\usepackage{amsmath,amssymb}
\usepackage{array,booktabs,tabularx,longtable}
\usepackage{calc}
\usepackage{graphicx}
\usepackage{flafter}
\usepackage{xcolor}
\usepackage{fancyhdr}
\usepackage{caption}
\usepackage{natbib}
\usepackage{fancyvrb}
\usepackage{seqsplit}
\usepackage{etoolbox}
\usepackage{newunicodechar}
\definecolor{bilipink}{RGB}{251,114,153}
\definecolor{biliblue}{RGB}{0,140,210}
\definecolor{shadecolor}{RGB}{248,248,248}
\IfFileExists{footnotehyper.sty}{\usepackage{footnotehyper}}{\usepackage{footnote}}
\makesavenoteenv{longtable}
\IfFileExists{xurl.sty}{\usepackage{xurl}}{}
\usepackage[
  colorlinks,
  citecolor=bilipink,
  linkcolor=bilipink,
  urlcolor=bilipink,
  breaklinks=true
]{hyperref}
\usepackage{bookmark}

\makeatletter
\patchcmd\longtable{\par}{\if@noskipsec\mbox{}\fi\par}{}{}
\makeatother

\DefineVerbatimEnvironment{Highlighting}{Verbatim}{commandchars=\\\{\},fontsize=\small}

\IfFileExists{framed.sty}{\usepackage{framed}}{%
  \newenvironment{snugshade}{}{}}

\newunicodechar{Δ}{\ensuremath{\Delta}}
\newunicodechar{κ}{\ensuremath{\kappa}}
\newunicodechar{∈}{\ensuremath{\in}}
\newunicodechar{−}{\ensuremath{-}}
\newunicodechar{♭}{\ensuremath{\flat}}

\makeatletter
\def\@maketitle{%
  \newpage
  \null
  \vspace{-3.5em}
  \begin{center}%
    {\color{biliblue}\hrule height 0.8pt}
    \vspace{1.5em}
    {\huge\bfseries\@title\par}
    \vspace{1.5em}
    {\color{biliblue}\hrule height 0.8pt}
    \vspace{1.6em}
    {\large\bfseries Yining Wang\par}
    \vspace{0.35em}
    {\normalsize Bilibili Inc.\par}
    \vspace{0.15em}
    {\small\href{mailto:10inspiral@gmail.com}{10inspiral@gmail.com}\par}
  \end{center}%
  \par
  \vspace{1.2em}}
\makeatother

\title{Beyond Prompted Agreement: Instruction Effects on Key and Beat Grouping in Text-to-Music Generation}
\author{Yining Wang}
\date{}

\hypersetup{
  pdftitle={Beyond Prompted Agreement: Instruction Effects on Key and Beat Grouping in Text-to-Music Generation},
  pdfauthor={Yining Wang},
  pdfsubject={Musical instruction following in text-to-music generation}
}

\begin{document}

\maketitle
\thispagestyle{fancy}

\begin{abstract}
In music-generation models, an explicit instruction to realize a musical attribute can paradoxically make it less common. How, then, should we assess whether a model follows a musical instruction? We use matched inputs that omit an attribute, specify one target, or specify another target, and compare the resulting audio. We evaluate key and beat grouping in ACE-Step 1.5, Stable Audio 3 and LeVo2 on 2,304 primary outputs. In Stable Audio 3, a four-beat instruction reduces detected four-beat occurrence from 96.9\% to 56.3\%, even though four beats remain more common than three under their respective instructions. Further experiments show that rephrasing can improve attainment while reducing the difference between target instructions, and that ACE-Step shows larger changes in target occurrence when text instructions retain its native attribute field. Our findings suggest that evaluating musical instruction following requires both target attainment and comparisons with generation with an omitted target instruction and with a different specified target.
\end{abstract}

\section{Introduction}
Text-to-music models allow users to describe musical attributes such as key and time signature. A listener who asks for a piece in a particular key expects the text instruction to influence the music. Finding the specified key in the output appears to support this expectation. However, the same key may also occur without the instruction. To what extent does matching a specified attribute reflect the influence of the instruction?

We investigate this question through global key and three- versus four-beat grouping. These attributes describe familiar aspects of tonal and rhythmic structure and have concrete target values. In Stable Audio 3, we find a surprising pattern: instructing the model to use four-beat grouping reduces detected four-beat occurrence from 96.9\% to 56.3\%, while instructing the model to use three-beat grouping increases detected three-beat occurrence from 1.6\% to 48.4\%. Four beats have higher target attainment, yet their occurrence decreases after the instruction. A small human listening assessment agrees with the direction of these changes.

\begin{samepage}
This observation motivates three questions:
\begin{enumerate}
\item \textbf{What does the model generate when no target is specified?}
\item \textbf{Does the output change when the specified target changes?}
\item \textbf{Are the results stable when the same instruction is phrased differently?}
\end{enumerate}
\end{samepage}

To answer these questions, we use a matched evaluation protocol. For each musical context, we generate audio without an instruction for the evaluated attribute and with each of two target instructions. The comparisons measure default occurrence, change after adding an instruction, and the effect of changing the specified target. We apply the protocol to ACE-Step 1.5, Stable Audio 3 Medium and LeVo2, using each system's supported input format.

Our study contributes experimental evidence that target attainment and changes in output under an instruction can diverge, together with a protocol for evaluating this difference. We further compare alternative instruction expressions and, for ACE-Step, text instructions with and without its native attribute field. These experiments show how the expression and input format affect the measured changes in target occurrence. We use the findings to identify the comparisons needed to assess musical instruction following.

\section{Related Work}
\paragraph{Music generation and musical attributes.}
Text-to-music models generate audio from descriptions of style, instruments and other musical properties \citep{copetSimpleControllableMusic2024,evansStableAudioOpen2024,yuanYuEScalingOpen2025}. Several systems provide more detailed control. FIGARO conditions symbolic generation on descriptions that include time signature and chords. MuseMorphose controls rhythmic intensity and polyphony, while MuseCoco and Text2midi generate symbolic music from textual descriptions \citep{rutteFIGAROGeneratingSymbolic2024,wuMuseMorphoseFullSongFineGrained2022,luMuseCocoGeneratingSymbolic2023,bhandariText2midiGeneratingSymbolic2025}. For audio generation, Mustango uses key, tempo, beat and chord information, and MusiConGen and Music ControlNet provide rhythm and harmony conditioning \citep{melechovskyMustangoControllableTexttoMusic2024,lanmusicongenrhythmandchordcontrolfortran,wumusiccontrolnetmultipletimevaryingcontro}. We study two of these common attributes, key and beat grouping, and assess how their occurrence changes when they are specified by an instruction.

\paragraph{Music understanding and evaluation.}
\citet{weiMusicGenerationModels2024} introduce SynTheory to probe whether music generation models encode music theory concepts. Their results suggest that these concepts can be recovered from internal representations, with performance varying across models and layers. Our study concerns how an instruction affects generated audio. In music evaluation, MusicEval measures overall musical impression and text alignment, SongEval assesses song aesthetics, and Music Arena collects listener preferences \citep{liuMusicEvalGenerativeMusic2025,yaoSongEvalBenchmarkDataset2025,kimMusicArenaLive2025}. \citet{grotschlaBenchmarkingMusicGeneration2025} compare automatic metrics with human judgments. These evaluations help assess the quality and alignment of generated music. Attribute-specific evaluation provides a further view: Mustango compares extracted keys and beat counts with the specified values \citep{melechovskyMustangoControllableTexttoMusic2024}. We retain this target-attainment measure and add comparisons with an omitted instruction and an alternative target.

\paragraph{Paired evaluation of instructions.}
Minimal-pair tests assess model behavior by varying selected parts of an input \citep{ribeiroBeyondAccuracyBehavioral2020}. \citet{yangComparedWhatBaselines2026} show that targeted text edits also change surface form, motivating comparisons with benign perturbations and paired analysis. We adopt paired comparisons for musical instructions and evaluate each complete input as received by the model. In audio, this also requires checking how reliably the specified attribute can be recognized. We compare automatic labels with expert judgments and use alternative expressions to assess whether the observed changes depend on the wording.

\section{Evaluating Musical Instruction Following}
We compare three inputs for the same musical context: one without the evaluated attribute and two specifying different values of that attribute. This design allows us to measure both what the model produces and how its output changes with the instruction.

\subsection{Matched inputs}
A matched family $f$ contains a musical context $c_f$, two targets $a_f,b_f$, and the generated outputs $y_f^0,y_f^a,y_f^b$. We hold the generator, random seed, duration and decoding settings fixed within the family. Each model's adapter converts the context and target into its supported input format. The complete family is the unit of analysis.

Table~\ref{tab:case-design} shows an example for Stable Audio 3. The three inputs share the genre, instrumentation and tempo, while the specified beat grouping changes. ACE-Step receives the target through a structured field, and LeVo2 receives it through a type-info prefix. Thus, our comparisons assess the behavior of each system through its available interface. The instruction expression and the specified target together form the input intervention.

\begin{table}[t]
\centering\small
\begin{tabular}{@{}>{\raggedright\arraybackslash}p{0.24\columnwidth}>{\raggedright\arraybackslash}p{0.68\columnwidth}@{}}
\toprule Condition & Stable Audio 3 prompt\\\midrule
No beat instruction & classical, piano + strings, 80 BPM\\[3pt]
3-beat instruction & classical, piano + strings, 80 BPM, \textbf{3/4 time signature}\\[3pt]
4-beat instruction & classical, piano + strings, 80 BPM, \textbf{4/4 time signature}\\
\bottomrule
\end{tabular}
\caption{A matched beat-grouping family. The added instruction is bold; musical context, seed, 30-second duration and decoding settings are shared.}\label{tab:case-design}
\end{table}

\subsection{Evaluation metrics}
Let $s(y,t)\in[0,1]$ measure how well output $y$ realizes target $t$, and let $u$ denote the other target in the same family. We average over the sampled families containing $t$:
\begin{align}
 N_t &= \mathbb E_f[s(y_f^0,t)],\\
 \operatorname{Acc}_t &= \mathbb E_f[s(y_f^t,t)],\\
 \Delta_t &= \mathbb E_f[s(y_f^t,t)-s(y_f^0,t)],\\
 R_t &= \mathbb E_f[s(y_f^t,t)-s(y_f^u,t)].
\end{align}
$N_t$ is the target's occurrence without an instruction for the evaluated attribute. Target attainment, $\operatorname{Acc}_t$, measures how often the specified value is detected. The gain $\Delta_t$ measures the effect of adding a target instruction, compared with generation without it. To measure the effect of changing the specified target, $R_t$ takes the difference in the same target's occurrence between its own instruction and the other target instruction.

For example, $R_4$ measures how much more often four-beat grouping occurs when the instruction specifies four rather than three beats. Both inputs contain a beat instruction, and only the specified target changes. By contrast, $\Delta_4$ compares the four-beat instruction with no beat instruction. We report each target separately to preserve differences between the specified values.

\subsection{Interpreting the comparisons}
The metrics address different questions. High attainment can arise when a target is already common without a target instruction. A positive gain indicates that adding the instruction increases its occurrence. A positive $R_t$ indicates that the specified target occurs more often under its own instruction than under the alternative instruction.

A model that generates four beats by default and correctly follows either beat instruction illustrates why these comparisons matter. It has perfect attainment and positive $R_t$ values for both targets, but zero four-beat gain: the four-beat instruction has nothing to increase. Thus, a small gain can accompany accurate target selection.

We use the metrics together to describe the observed behavior. The comparison with an omitted target instruction includes the added instruction expression, while the comparison between target instructions holds that expression fixed and changes the specified value. Both are defined for the selected musical contexts and input formats. Appendix~\ref{app:algebra} relates these target-specific metrics to the aggregate margin and describes the special case of complementary binary scores.

\section{Experimental Design}
We evaluate key and beat grouping across three music generation systems. We vary the musical context to explore whether the instruction effects persist across genres, instruments and tempos.

\subsection{Tasks and systems}
Global key describes tonal organization through 12 major and 12 minor tonic--mode classes. We use exact agreement with the specified key as the primary score and related-key MIREX credit as a secondary measure \citep{raffelMirEvalTransparent2014}. Beat grouping describes rhythmic organization through the number of beats between detected downbeats. We evaluate targets of three and four beats. Instructions use the systems' supported time-signature inputs, while scoring measures the beat grouping found in the audio.

We select systems with different generation architectures. ACE-Step 1.5 combines a language-model planner with a diffusion Transformer (DiT) acoustic renderer \citep{gongACEStep15Pushing2026}. Stable Audio 3 Medium uses a text-conditioned latent DiT \citep{evansStableAudio32026}. LeVo2 combines hierarchical autoregressive token modeling with diffusion-based audio decoding \citep{leiLeVo2Stable2026}. Each system generates 30-second clips. Key and beat instructions enter through structured fields in ACE-Step, prompt text in Stable Audio 3, and a type-info prefix in LeVo2. These input formats are fixed for the primary evaluation.

\subsection{Musical contexts and generation}
We construct contexts from four broad genres and vary instrumentation and BPM within each genre. For key, each of the 24 keys appears four times as a specified target in each genre. This gives $24\times4\times4=384$ target appearances, paired into 192 families. Same-tonic major/minor pairs test mode changes, and distant-tonic pairs test changes of tonic.

For beat grouping, we cross four instrumentation palettes with four BPM values in each genre, giving 64 families. We exclude descriptions that directly suggest a target, such as \emph{waltz}, and contexts that provide insufficient tonal or rhythmic information for the task. The two tasks comprise 256 families and 768 primary outputs per system, or 2,304 outputs in total. Case manifests specify the exact contexts and target pairs.

We generate all three conditions of a family with the same settings. This preserves the pairing when comparing their outputs. Appendix~\ref{app:provenance} records the input formats, model versions, generation settings and accepted scoring caches. Additional repeated-generation checks are reported in Appendix~\ref{app:seed}.

\subsection{Audio scoring and validation}\label{sec:scoring}
\begin{figure}[t]
\centering
\includegraphics[width=\columnwidth]{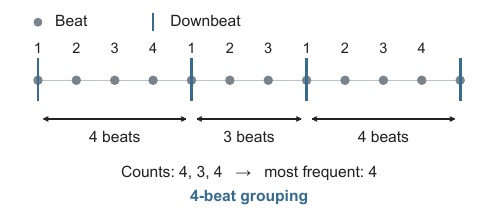}
\caption{Schematic example: interval counts of 4, 3 and 4 yield a four-beat grouping. The starting downbeat is counted; the next belongs to the following interval.}\label{fig:beat-counting}
\end{figure}

We estimate key with S-KEY \citep{kongSKEYSelfsupervisedLearning2025} and detect beats and downbeats with Beat This \citep{foscarinBeatThisAccurate2024}. To obtain beat grouping, we count beats in each complete interval between successive downbeats and take the most frequent count (Figure~\ref{fig:beat-counting}). We skip intervals containing no detected beats and resolve ties using the count that occurs first. Labels other than three or four score zero for both targets. Clips with insufficient detected events are marked undetermined. Failed generations and undetermined clips remain in the denominator for attainment.

We assess recognition on generated music using expert judgments. Five experts participate without seeing the instructions or automatic labels. In a 60-clip assessment with three ratings per clip, reweighted agreement with the clear expert majority is 0.796 for exact key and 0.745 for beat grouping. These checks provide evidence for using the tools on generated audio. A separate assessment of five complete Stable Audio 3 beat families by one annotator gives gains of $-0.2$ for four-beat instructions and 0.4 for three-beat instructions, agreeing in direction with the primary automatic results. Appendix~\ref{app:seed} describes the sampling and weighting of the assessments.

We compute the main tables and output distributions from the same recognition results, using the same scoring rules for all systems. We estimate uncertainty by resampling complete families and retaining the three outputs together. Full numerical intervals and the resampling procedure are reported in Appendix~\ref{app:provenance}.

\section{Results and Discussion}
Overall, key instructions and beat instructions affect target occurrence differently. ACE-Step and Stable Audio 3 show clear changes in key occurrence. For beat grouping, attainment, gain and the effect of changing the target describe different aspects of the same behavior.

\subsection{Key}
ACE-Step and Stable Audio 3 increase the occurrence of a specified key and shift detected keys when the target changes (Table~\ref{tab:key}). Their exact-key attainment is similar, at 66.7\% and 67.4\%, respectively. Thus, the instructions clearly change key occurrence in both systems, although the recognizer detects the exact specified key in only about two thirds of outputs.

LeVo2 shows much smaller changes through its evaluated interface. Its key attainment is low, and its mean $R_t$ is small and uncertain. These differences show why attainment and instruction effects should be reported together: they describe both the frequency of the desired result and the change in output when the input changes. Appendix~\ref{app:key} reports estimates for individual keys.

\begin{table*}[t]
\centering\small
\begin{tabular}{lrrrr}
\toprule System & Neutral & Acc & $\Delta$ & Mean $R$\\\midrule
ACE-Step & 0.055 & 0.667 & 0.612 & 0.628 \\
LeVo2 & 0.029 & 0.055 & 0.026 & 0.021 \\
Stable Audio 3 & 0.029 & 0.674 & 0.646 & 0.635
\\
\bottomrule
\end{tabular}
\caption{Global-key results over 192 complete families per system. Each family contributes its two target appearances. This table and the primary beat tables share the scoring policy described in Section~\ref{sec:scoring}.}\label{tab:key}
\end{table*}

\subsection{Beat grouping}
\begin{figure*}[t]
\centering
\includegraphics[width=\textwidth]{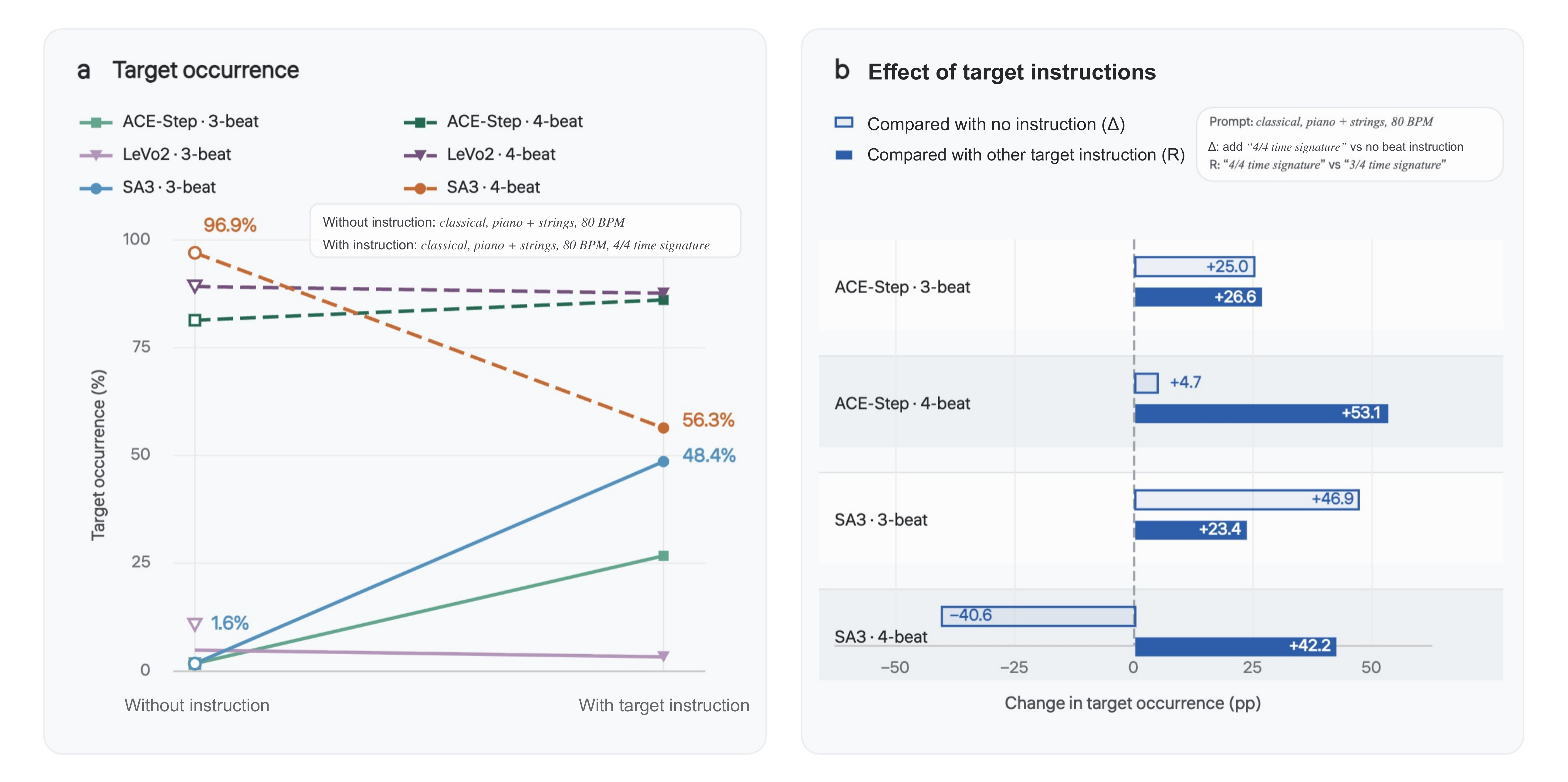}
\caption{Beat-grouping attainment and instruction effects.
(a) Target occurrence without and with the corresponding target instruction.
(b) Changes in target occurrence for ACE-Step and Stable Audio 3, relative to no instruction ($\Delta_t$, outlined bars) and the other target instruction ($R_t$, filled bars). Both comparisons share a zero line; values are percentage points. Stable Audio 3's four-beat occurrence decreases by 40.6 points relative to no instruction, yet increases by 42.2 points relative to a three-beat instruction. LeVo2 is omitted from (b) because its beat-instruction contrasts are small under the evaluated interface (Table~\ref{tab:beat}); its occurrence rates remain in (a). Solid and dashed lines in (a) denote three- and four-beat targets. Insets give Stable Audio 3 instruction examples.
Estimates are means over 64 complete families per interface, with one seed per family.}
\label{fig:primary-response}
\end{figure*}

The beat-grouping results reveal a different pattern (Table~\ref{tab:beat}, Figure~\ref{fig:primary-response}). For three beats, ACE-Step and Stable Audio 3 have positive gains and $R_t$ values. Their attainment rates are 26.6\% and 48.4\%, respectively. Both models produce more three-beat outputs when three beats are specified, though their attainment differs substantially.

For four beats, the comparison with no instruction becomes particularly informative. Stable Audio 3 produces four-beat grouping in 96.9\% of neutral outputs but only 56.3\% of outputs under a four-beat instruction. An explicit instruction to generate the target therefore makes it less common. Nevertheless, four-beat occurrence is 42.2 percentage points higher under a four-beat instruction than under a three-beat instruction. Changing the specified target changes four-beat occurrence, while both instructions move the output away from its neutral distribution.

ACE-Step has a small positive four-beat gain and a larger difference between target instructions, with $R_4=0.531$. Its four-beat instruction preserves every neutral clip detected as four-beat. A three-beat instruction produces a more dispersed distribution, increasing three-beat occurrence while leaving four-beat outputs more frequent than three-beat outputs.

Stable Audio 3 redistributes outputs under both instructions. Three-beat instructions make three beats its most frequent detected label. Under four-beat instructions, 26 initially four-beat clips change label: 16 become three-beat and ten receive other grouping labels. LeVo2, in contrast, remains predominantly four-beat across all inputs, with 56--58 of 64 clips receiving that label in each condition. Its gains and differences between target instructions are small. Appendix~\ref{app:distribution} gives the complete distributions for all three systems.

\begin{table*}[t]
\centering\small
\begin{tabular}{llrrrr}
\toprule System & Target beat grouping & Neutral & Acc & $\Delta$ & $R_t$\\\midrule
ACE-Step & 3 & 0.016 & 0.266 & 0.250 & 0.266 \\
ACE-Step & 4 & 0.813 & 0.859 & 0.047 & 0.531 \\
LeVo2 & 3 & 0.047 & 0.031 & -0.016 & 0.000 \\
LeVo2 & 4 & 0.891 & 0.875 & -0.016 & -0.031 \\
Stable Audio 3 & 3 & 0.016 & 0.484 & 0.469 & 0.234 \\
Stable Audio 3 & 4 & 0.969 & 0.563 & -0.406 & 0.422
\\
\bottomrule
\end{tabular}
\caption{Target-specific beat-grouping results, 64 complete families per system. Targets 3 and 4 denote three and four beats per detected downbeat interval, not verification of a notated time-signature denominator.}\label{tab:beat}
\end{table*}

\subsection{Off-attribute controls}
We also measure beat grouping in audio generated under key instructions, and key in audio generated under beat instructions. These comparisons assess changes under instructions directed at a different musical attribute. For ACE-Step and Stable Audio 3, unrelated key instructions yield three-beat changes of 0.003--0.008, compared with 0.250 and 0.469 under the primary beat instructions. Stable Audio 3's four-beat change is $-0.060$ under key instructions, smaller than its change under beat instructions. Thus, the measured beat changes are larger for beat-directed instructions. The two tasks use different contexts and instruction expressions, so these comparisons concern the overall instruction types.

\subsection{Instruction expressions}
To assess sensitivity to wording, we replace the original fractional time-signature instructions with two verbal expressions on 16 Stable Audio 3 families. We use three random seeds to generate 192 new clips and score them alongside the original clips for all three conditions in the same environment. The expressions are \emph{three/four beats per bar} and \emph{The music has three/four beats in each bar.}

\begin{figure*}[t]
\centering
\includegraphics[width=\textwidth]{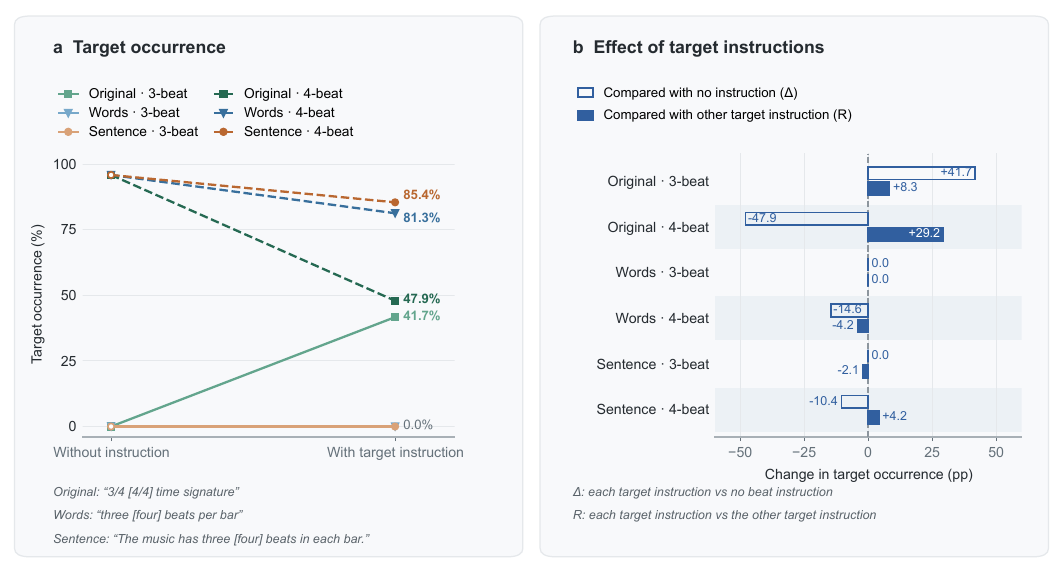}
\caption{Beat-grouping attainment and instruction effects under three instruction expressions in Stable Audio 3.
(a) Target occurrence without and with the corresponding target instruction; solid and dashed lines denote three- and four-beat targets.
(b) Changes relative to no instruction ($\Delta_t$, outlined bars) and the other target instruction ($R_t$, filled bars), sharing a zero line in percentage points. Rephrasing increases four-beat attainment, but occurrence remains below the neutral baseline and differences between target instructions are near zero. Observed three-beat attainment is zero under both alternatives.
Estimates are means over the same 16 complete families and three seeds, averaged within family. The reused neutral baseline is 0\% for three beats and 95.8\% for four beats, separate from the 64-family primary experiment. Italic examples give the instruction expressions; Appendix~\ref{app:seed} reports complete numerical intervals.}
\label{fig:wording-sensitivity}
\end{figure*}

We observe that both alternatives increase four-beat attainment while removing detected three-beat attainment (Figure~\ref{fig:wording-sensitivity}). Three-beat attainment falls from 41.7\% to zero. Four-beat attainment rises from 47.9\% to 81.3\% and 85.4\%, although both remain below the 95.8\% neutral baseline for these families.

The higher four-beat attainment is accompanied by little change in target occurrence when the specified target changes. Under either expression, outputs tend to be four-beat regardless of whether three or four beats are specified. The four-beat $R_t$ values are $-0.042$ and 0.042, with both intervals including zero. These observations suggest that an expression can increase target attainment while weakening the distinction between the two instructions.

\subsection{Input format and model differences}
An independent experiment compares the original instructions with a full time-signature sentence and the expression \emph{Use three/four beats per bar}. It uses the same 16 contexts with one random seed and comprises 256 Stable Audio 3/LeVo2 outputs and 192 new ACE-Step outputs. The alternatives and generation settings are defined separately from the preceding wording experiment.

Stable Audio 3 retains three-beat attainment with the full sentence (8/16), but produces none with the \emph{Use three beats per bar} instruction. LeVo2 remains predominantly four-beat, with three-beat attainment of 0/16, 1/16 and 0/16 for the original, sentence and words conditions. Thus, the changes in target occurrence vary across both expressions and systems.

For ACE-Step, we compare text alone with text accompanied by the native time-signature field. Three-beat attainment is 4/16 under the native field or sentence alone, and zero under words alone. The four-beat $R_t$ value falls from 0.375 under the native field to 0.125 and 0.063 under sentence-only and words-only inputs. Keeping the field alongside the same text yields $R_4$ values of 0.625 and 0.563. These comparisons show the importance of the input format: replacing a structured field with text changes how the instruction enters the model as well as how it is expressed. Appendix~\ref{app:wording} gives the independent baselines, field-retained controls and full results.

\subsection{Implications for evaluating instruction following}
The results illustrate three aspects of musical instruction following that benefit from separate evaluation.

\textbf{Default occurrence.} A specified attribute may already be common in generated music. Stable Audio 3's four-beat result shows that a relatively high attainment rate can accompany a decrease from generation without a target instruction. Measuring the default occurrence makes this decrease visible.

\textbf{Changing the specified target.} Changing the specified target can shift the output even when adding an instruction reduces its occurrence relative to the default. The positive four-beat $R_t$ therefore describes a different comparison from its negative gain. Reporting both targets also reveals differences that an aggregate score can conceal.

\textbf{Expression sensitivity.} The wording experiments show that improved attainment can accompany a smaller difference between target instructions. The independent experiment further shows that this pattern varies with expression and input format. To assess instruction following in practice, we need to examine whether the system continues to distinguish specified targets across the expressions and formats available to users.

These observations motivate evaluating default output, target attainment and the effect of changing the specified target together. The comparisons can also be constructed for attributes such as emotion, timbre, instrumentation and local musical structure. Future work could extend the target sets and pair automatic recognition with human judgments of target realization, rhythmic clarity and audio quality.

\section{Conclusion}
We evaluate musical instruction following through matched inputs with omitted and alternative-target instructions. Our experiments on key and beat grouping show that target attainment and changes in output under an instruction can diverge. Stable Audio 3 provides a clear example: instructing the model to use four-beat grouping reduces its occurrence, while alternative expressions can improve attainment and reduce the difference between target instructions. The changes in ACE-Step's target occurrence also depend on whether textual instructions retain its native field. These findings suggest that evaluating instruction following requires measuring what a model generates by default and how its output changes with the specified target and expression.

\section*{Limitations}
The evaluation covers two global attributes, three systems and a restricted set of genres, instruments and tempos. Results describe the selected contexts and each system's available interface. Model architecture, training data and input format vary together across systems.

The audio recognizers have domain-dependent errors. The expert assessments provide evidence about agreement and the direction of selected effects, but their small model-specific samples limit validation of individual targets and system rankings. Larger assessments should distinguish target realization from rhythmic ambiguity and audio degradation.

The wording experiments use 16 families and change several lexical and syntactic features together. Their results concern the tested expressions rather than individual words. ACE-Step's text-only comparisons also change the input channel, which we examine using field-retained controls. Its newly generated original conditions use a separate runtime baseline because they differ from historical PCM. The two wording experiments retain their own baselines and are analyzed separately. Repeated-generation checks and complete uncertainty estimates are provided in the appendix.

\section*{Ethical Considerations}
This study evaluates released music generation systems through controlled instructions about musical attributes. We consider its additional ethical, societal and environmental risks to be limited, with the main considerations being participant privacy and the distribution of audio and model assets.

Participants confirmed their informed consent via written messages for all listening assessments. Assessment findings are reported in aggregate. Supplementary rating files use anonymous rater identifiers and omit names, contact details and free-text comments; original rating forms and organizer mappings are excluded. The case design concerns musical attributes and does not infer artist identity or listener demographics.

The supplementary code and data packages include source-specific license notices and attribution, as detailed in Appendix~\ref{app:external} and their artifact documentation. Model weights and source audio are obtained from their providers. External reference labels are included under verified annotation licenses or supplied through acquisition links.

\appendix
\section{Metric Identities and Interpretation}\label{app:algebra}
A symmetric family summary is the margin,
\begin{align}
 M_f=\tfrac12[&s(y_f^a,a)-s(y_f^a,b)\nonumber\\
              +&s(y_f^b,b)-s(y_f^b,a)].
\end{align}
The margin is the mean of the two $R_t$ values, each measuring the effect of changing the specified target.

For a controller that always generates four beats without an instruction and correctly realizes either specified target, $\operatorname{Acc}_3=\operatorname{Acc}_4=1$, $\Delta_3=1$, $\Delta_4=0$, and $R_3=R_4=1$. The zero four-beat gain reflects perfect default attainment.

For one family, write $s_{ij}=s(y^i,j)$. Then
\begin{align}
R_a+R_b&=s_{aa}-s_{ba}+s_{bb}-s_{ab}\\
       &=2M.
\end{align}
This identity holds without binary outputs. In contrast, the within-output target contrast $s_{aa}-s_{ab}$ differs from $R_a=s_{aa}-s_{ba}$ in general. Averaging the two within-output contrasts happens to recover $M$; it does not make them interchangeable for each specified target.

If $s(y,a)+s(y,b)=1$ for every output, let $p_i=s(y^i,a)$. Then
\begin{align}
R_a&=p_a-p_b,\\
R_b&=(1-p_b)-(1-p_a)=p_a-p_b,\\
\Delta_a+\Delta_b&=(p_a-p_0)+(p_0-p_b)=p_a-p_b.
\end{align}
Thus $M=R_a=R_b=\Delta_a+\Delta_b=2\bar\Delta$. Writing $\overline{\operatorname{Acc}}=\tfrac12[s(y^a,a)+s(y^b,b)]$, the same complement assumption gives $\bar\Delta=\overline{\operatorname{Acc}}-\tfrac12$ and $M=2\overline{\operatorname{Acc}}-1$. No independence claim is warranted. With other labels, soft noncomplementary scores, or failures scored zero for both targets, the last set of equalities need not hold.

\section{Cases, Interfaces and Reproducibility}\label{app:provenance}
Each case records family identity, condition role, targets, genre, instrumentation, BPM, duration and seed. Four genres each have four palettes. Classical palettes are piano with strings, string ensemble, piano trio and small orchestra. Pop uses piano, guitar or keys with bass and drums, or a full pop band. Jazz uses piano trio, guitar trio, piano quartet or vibraphone trio. Rock uses guitar, dual guitars, keyboard band or piano with bass and drums. Case manifests, rather than an implied full Cartesian product for key, specify the exact assignments.

For key, same-tonic major/minor pairs appear in both orientations. Additional within-mode tonic pairs avoid relative, parallel and adjacent-fifth matches. Each of the 24 targets appears twice in each orientation per genre. The resulting 384 target appearances correspond to 192 two-target families. Beat grouping has one pair crossed with four genres, four palettes and four BPM values, giving 64 families.

The family seed is
\[
\operatorname{int}(\operatorname{SHA256}(f\mathbin{:}r),16)\bmod 2^{31},
\]
where $r$ is the replicate index. All three conditions share it; random states are reset before each generation. A seed is a pairing device, not a common latent representation across systems.

\paragraph{Frozen generation settings.}
ACE-Step uses the released DiT \texttt{acestep-v15-xl-sft} and LM \texttt{acestep-5Hz-lm-4B}: eight steps, DiT guidance 7.0, Euler ODE, LM temperature 0.85, CFG 2.0, top-$p$ 0.9, thinking enabled, batch one, instrumental output at 48 kHz. The archived ACE-Step run records the model names and generation parameters, but not the generation code commit identifier or checkpoint file checksums. Stable Audio 3 uses revision \texttt{5866e5b415a2}, eight steps, CFG 1.0, half precision, batch one and chunked decoding, at 44.1 kHz. LeVo2 uses SongGeneration-v2-large, revision \texttt{653cbcf47161}, BGM generation, CFG 1.5, temperature 0.9, top-$k$ 50, top-$p$ zero, stride five and chunked decoding, at 48 kHz.

ACE-Step passes key and grouping in native fields (Unicode flat spellings for key); neutral leaves the scored field empty. Stable Audio 3 appends the target to the context text. LeVo2 replaces instrumental ``+'' separators with commas, prepends \texttt{[Musicality-very-high], [Pure-Music]}, and sets lyrics to a period. The compatibility adapter restores the official type tokens and checks the embedding size against the checkpoint. These choices are fixed parts of the evaluated interface.

\paragraph{Scoring window and cache policy.}
Evaluation begins at the detected musical onset and uses up to 30 seconds, without repetition or padding. A remainder below 15 seconds is invalid. The primary analysis uses the accepted key caches and the complete corrected-runtime beat cache for each model. Four LeVo2 beat labels change; the revised neutral and four-beat-instruction counts are 57/64 and 56/64, rather than the historical 58/64 and 55/64. The revised $\Delta_4$ is $-1/64$ rather than $-3/64$. This difference is a measurement correction, not a new generator result. Both cache versions and their hashes are retained.

\paragraph{Uncertainty and primary estimates.}
We use 20,000 genre-stratified bootstrap draws over complete families, preserving the pairing of the three outputs in each family. Numerical intervals in this appendix are exploratory pointwise 95\% intervals, shown after the estimates. A supplementary machine-readable table also gives Bonferroni bounds for the 12 primary beat $\Delta/R$ contrasts. These post-review analyses were not preregistered. The replication subset jointly resamples observed seeds and families. Tables~\ref{tab:key-intervals} and~\ref{tab:beat-intervals} give the full primary results. Stable Audio 3's four-beat change under unrelated key instructions is $-0.060$ [$-0.109,-0.013$].
\begin{table*}[t]
\centering\small
\begin{tabular}{lrrrr}
\toprule System & Neutral & Acc & $\Delta$ & Mean $R$\\\midrule
ACE-Step & 0.055 & 0.667 [0.617, 0.716] & 0.612 [0.563, 0.661] & 0.628 [0.568, 0.685] \\
LeVo2 & 0.029 & 0.055 [0.034, 0.078] & 0.026 [0.003, 0.049] & 0.021 [-0.003, 0.047] \\
Stable Audio 3 & 0.029 & 0.674 [0.628, 0.721] & 0.646 [0.599, 0.693] & 0.635 [0.581, 0.688]
\\
\bottomrule
\end{tabular}
\caption{Primary key estimates with 95\% intervals, 192 complete families per system.}\label{tab:key-intervals}
\end{table*}
\begin{table*}[t]
\centering\small
\begin{tabular}{llrrrr}
\toprule System & \shortstack{Target beat\\grouping} & Neutral & Acc & $\Delta$ & $R_t$\\\midrule
ACE-Step & 3 & 0.016 & 0.266 [0.172, 0.359] & 0.250 [0.156, 0.344] & 0.266 [0.172, 0.359] \\
ACE-Step & 4 & 0.813 & 0.859 [0.766, 0.938] & 0.047 [0.000, 0.109] & 0.531 [0.406, 0.641] \\
LeVo2 & 3 & 0.047 & 0.031 [0.000, 0.078] & -0.016 [-0.063, 0.031] & 0.000 [-0.047, 0.047] \\
LeVo2 & 4 & 0.891 & 0.875 [0.813, 0.938] & -0.016 [-0.078, 0.047] & -0.031 [-0.109, 0.047] \\
Stable Audio 3 & 3 & 0.016 & 0.484 [0.375, 0.594] & 0.469 [0.344, 0.594] & 0.234 [0.078, 0.391] \\
Stable Audio 3 & 4 & 0.969 & 0.563 [0.469, 0.656] & -0.406 [-0.516, -0.297] & 0.422 [0.297, 0.547]
\\
\bottomrule
\end{tabular}
\caption{Primary beat-grouping estimates with 95\% intervals, 64 complete families per system.}\label{tab:beat-intervals}
\end{table*}

\section{Complete Beat Distributions}\label{app:distribution}
Table~\ref{tab:dist} gives counts under the primary corrected scoring policy. ``Other'' includes positive grouping counts outside three/four. No clip in this primary table is undetermined, but that category is retained explicitly rather than removed by filtering.
\begin{table*}[t]
\centering\small
\begin{tabular}{llrrrrl}
\toprule System & Instruction & 3 & 4 & Other & Undetermined & Other counts \\\midrule
ACE-Step & Neutral & 1 & 52 & 11 & 0 & 1:6, 2:5 \\
ACE-Step & Three & 17 & 21 & 26 & 0 & 1:12, 2:6, 5:1, 6:7 \\
ACE-Step & Four & 0 & 55 & 9 & 0 & 1:3, 2:6 \\
Stable Audio 3 & Neutral & 1 & 62 & 1 & 0 & 2:1 \\
Stable Audio 3 & Three & 31 & 9 & 24 & 0 & 2:7, 5:2, 6:12, 7:1, 12:2 \\
Stable Audio 3 & Four & 16 & 36 & 12 & 0 & 1:4, 2:5, 6:3 \\
LeVo2 & Neutral & 3 & 57 & 4 & 0 & 2:3, 6:1 \\
LeVo2 & Three & 2 & 58 & 4 & 0 & 1:2, 6:2 \\
LeVo2 & Four & 2 & 56 & 6 & 0 & 2:4, 6:2
\\
\bottomrule
\end{tabular}
\caption{Each row contains 64 outputs. In the final column, grouping:count reports the uncollapsed labels. These are recognizer outputs, not independently verified meter annotations.}\label{tab:dist}
\end{table*}

\section{External Reference Checks}\label{app:external}
\paragraph{Artifact distribution.}
The review packages retain MIT notices for ACE-Step and Stable Audio code and the academic, research and education-only terms for SongGeneration-related code. GTZAN key references retain MIT; RWC/AIST-derived grouping references retain the official annotation archive's CC BY-NC 4.0 license, attribution and change notices. Source audio, model weights and original annotation archives are excluded. For GiantSteps, Ballroom and GTZAN-Rhythm, the packages retain case identifiers, predictions and computed scores and provide acquisition links for the reference labels. Full notices, sources and review-use terms are supplied in \texttt{ARTIFACT\_LICENSES.md}, \texttt{TERMS\_OF\_USE.md} and \texttt{licenses/} in each package.

Table~\ref{tab:external} uses GiantSteps key annotations \citep{kneesGiantStepsTwoDatasets2015} with audio windows from CMI-Bench \citep{maCMIBenchComprehensiveBenchmark2025}, and Ballroom audio \citep{gouyonBallroomEvaluationRhythm2004} with grouping references mapped from the dance-style labels. RWC Classical recordings \citep{gotoRWCMusicDatabase2002} use AIST beat and downbeat annotations \citep{gotoAISTAnnotationRWC2006}. The two GTZAN checks use the original audio collection \citep{tzanetakisMusicalGenreClassification2002}, with version 1.0 key labels \citep{lerchGTZANKeyAnnotations2022} or GTZAN-Rhythm v2 annotations \citep{marchandGTZANRhythmExtending2015}.

\begin{table*}[t]
\centering\small
\begin{tabular}{llrl}
\toprule Task & Reference & $N$ & Estimate\\\midrule
Key & GiantSteps via CMI-Bench & 2,406 & Exact 0.506 [0.468,0.544]; MIREX 0.626 [0.594,0.658]\\
Key & GTZAN Keys, fault-filtered & 769 & Exact 0.656 [0.619,0.693]; MIREX 0.765 [0.739,0.790]\\
Grouping & Ballroom, excluding Samba & 602 & Balanced accuracy 0.998 [0.994,1.000]\\
Grouping & RWC Classical with AIST & 37 & Balanced accuracy 0.734 [0.599,0.867]\\
Grouping & GTZAN-Rhythm v2, fault-filtered & 619 & Balanced accuracy 0.806 [0.678,0.926]\\\bottomrule
\end{tabular}
\caption{Historical external validation with frozen task definitions. GiantSteps observations are windows and uncertainty is track-clustered. These reference estimates are not transferred as known accuracies on generated outputs.}\label{tab:external}
\end{table*}
Unknown GTZAN key labels are excluded. Both GTZAN checks use the published filtered lists from Corey Kereliuk's \texttt{dnn-mgr} repository,\footnote{\url{https://github.com/coreyker/dnn-mgr/tree/master/gtzan}} addressing repetitions and distortions documented by \citet{sturmGTZANDatasetContents2013}. RWC and GTZAN-Rhythm grouping references require at least 90\% of complete annotated bars to have the modal three/four grouping. Beat/downbeat F-measures are 0.937/0.863 for RWC and 0.918/0.831 for GTZAN-Rhythm. Metrical multiples, notably four-to-two, are an important error mode. Synthetic key cadences, metrical probes and silence checks supplement real-music validation. No external accuracy bound justifies assuming errors are invariant to a generation prompt.

\section{Repeated Generation and Human Assessment}\label{app:seed}
\paragraph{Repeated generation.}
To assess sensitivity to random seeds, we select 32 key and 16 beat families per model using their metadata and repeat generation with three random seeds.
\begin{table*}[!t]
\centering\footnotesize
\setlength{\tabcolsep}{2pt}
\begin{tabular}{@{}llrr@{}}
\toprule Model & Target & $\Delta$ & $R$\\\midrule
ACE & 3 & 0.208 [0.062,0.354] & 0.229 [0.083,0.375]\\
& 4 & 0.042 [-0.062,0.146] & 0.542 [0.354,0.729]\\
SA3 & 3 & 0.417 [0.250,0.583] & 0.083 [-0.188,0.312]\\
& 4 & -0.479 [-0.688,-0.271] & 0.292 [0.062,0.521]\\
LeVo2 & 3 & -0.021 [-0.083,0.000] & 0.000 [0.000,0.000]\\
& 4 & 0.021 [-0.125,0.188] & 0.062 [-0.104,0.250]\\\bottomrule
\end{tabular}
\caption{Three-seed replication subset, 16 beat families per model. ACE and SA3 denote ACE-Step and Stable Audio 3. Intervals jointly resample complete genre-stratified families and the three observed replicate indices. Degenerate bootstrap intervals reflect the observed sample only.}\label{tab:seed}
\end{table*}
The replication subset is metadata-selected, with four beat families per genre spanning the palettes and BPMs. Key has eight per genre and per pair slot. It is not used to increase primary sample size. One LeVo2 replicate-2 treatment has less than 15 seconds and is retained as a failed output. The post-review target-specific analysis uses 20,000 draws; numerical endpoints can differ slightly from earlier 10,000-draw aggregate tables.

For Stable Audio 3, repeating generation preserves positive three-beat gain (0.417) and negative four-beat gain ($-0.479$). Its three-beat $R_t$ value is small and uncertain (0.083), whereas its four-beat value remains positive (0.292). The direction of the gains is therefore more consistent than the differences between target instructions. Table~\ref{tab:seed} reports results and uncertainty for each target across all systems.

The 60-clip expert assessment uses three independent ratings per clip. The original assessment samples recognizer successes and failures equally, so agreement is reweighted to the primary-experiment population and computed on clips with a clear expert majority; the assessment report also records ambiguous judgments and agreement over all assessed clips.

The historical complete-triplet human audit covers 28 complete families, 14 per task, rated by one annotator without seeing the generation instructions or automatic labels. The pooled beat $\Delta$ is 0.036 for both human and automatic labels. By target, human versus automatic $\Delta_3$ is 0.143 versus 0.000, and $\Delta_4$ is $-0.071$ versus 0.071. On beat grouping, human $R_3=0.000$ [$-0.143,0.214$] and $R_4=0.357$ [0.071,0.643]; corresponding automatic estimates are 0.071 [$-0.143,0.286$] and 0.214 [0.000,0.500]. Its Stable Audio 3 subset has only five families. Neutral, three-beat-instruction and four-beat-instruction human--automatic agreement is respectively 4/5, 1/5 and 4/5. These condition differences reinforce why overall agreement cannot certify the main effect. The tiny model-specific cells are descriptive and are not assigned bootstrap precision when a genre stratum contains only one family.

\paragraph{Additional wording settings and default comparison.}
Table~\ref{tab:wording-three-seed} gives the numerical results plotted in Figure~\ref{fig:wording-sensitivity}. Its exploratory intervals use 10,000 complete-family bootstrap draws, averaging the three observed seeds within each family and retaining them together. Neutral occurrence is zero for three beats and 0.958 for four beats. This figure analysis is separate from the joint seed--family resampling in Table~\ref{tab:seed}; the independent one-seed experiment is also reported separately.
The alternatives append \emph{three/four beats per bar} or \emph{The music has three/four beats in each bar.} They differ from the original fractional time-signature expression in more than token length and are analyzed as complete wording interventions. Four-beat occurrence after rephrasing remains below the model's default level: without a beat-grouping instruction, 46/48 clips are detected as four-beat; the gains over neutral for the two expressions remain $-0.146$ and $-0.104$. Rephrasing thus partially restores default four-beat output without exceeding it.

\clearpage
\begin{table}[htbp]
\centering\small
\begin{tabular}{llrrr}
\toprule Form & Target & Acc & $\Delta_t$ & $R_t$\\\midrule
Original & 3 & 0.417 [0.229,0.604] & 0.417 [0.229,0.604] & 0.083 [-0.125,0.292]\\
Original & 4 & 0.479 [0.271,0.688] & -0.479 [-0.667,-0.292] & 0.292 [0.083,0.479]\\
Words & 3 & 0.000 [0.000,0.000] & 0.000 [0.000,0.000] & 0.000 [0.000,0.000]\\
Words & 4 & 0.813 [0.667,0.938] & -0.146 [-0.313,0.000] & -0.042 [-0.208,0.104]\\
Sentence & 3 & 0.000 [0.000,0.000] & 0.000 [0.000,0.000] & -0.021 [-0.063,0.000]\\
Sentence & 4 & 0.854 [0.729,0.958] & -0.104 [-0.208,-0.021] & 0.042 [-0.042,0.146]
\\
\bottomrule
\end{tabular}
\caption{Stable Audio 3 three-seed wording experiment, 16 families. Estimates followed by 95\% intervals. Zero-width empirical intervals reflect the observed sample and do not establish zero population uncertainty.}\label{tab:wording-three-seed}
\end{table}
\section{Independent Three-Model Expression Experiment}\label{app:wording}
This experiment fixes replicate zero of the same 16 beat families. Stable Audio 3 and LeVo2 each regenerate eight conditions: neutral; original three/four instructions; \emph{The time signature is 3/4.} or \emph{The time signature is 4/4.}; \emph{Use three beats per bar.} or \emph{Use four beats per bar.}; and \emph{Use a clear time signature.} Clauses are appended to the native neutral input. The target-free attribute mention is analyzed separately, not as additional neutral samples. Unlike Figure~\ref{fig:wording-sensitivity}, this experiment uses one seed and different alternatives; the two studies are not pooled.

ACE-Step adds four controls to the eight corresponding conditions, yielding 12 conditions per family. Text-only conditions leave its native time-signature field empty; the four additional controls retain the specified field while adding the same text. Native-to-text comparisons therefore change the input channel and cannot be interpreted as isolated lexical effects. ACE-Step uses \texttt{acestep-v15-xl-sft} and \texttt{acestep-5Hz-lm-4B}, generating 30-second, 48 kHz stereo instrumentals with 8 Euler ODE steps, guidance 7, LM temperature 0.85/CFG 2/top-p 0.9, thinking enabled and DCW disabled.

All 448 generated and scored identities passed acceptance: 128 each for Stable Audio 3/LeVo2 and 192 for ACE-Step. Checks cover prompts, seeds, time-signature fields, complete families, audio properties and hashes, and frozen code/model hashes. Both earlier models' 48 original-condition clips match historical PCM. ACE-Step originals do not reproduce historical PCM, so this fresh runtime baseline remains separate from the primary experiment; historical ACE code/checkpoint hashes were not archived. All outputs use the same frozen Beat This checkpoint and analysis window. Two remote analyses and the local replay agree byte-for-byte.

Table~\ref{tab:wording-two-model} reports attainment and paired effects for the eight-condition design. Neutral three/four occurrence is 0/16 and 16/16 in Stable Audio 3/LeVo2, and 2/16 and 13/16 in ACE-Step. Stable Audio 3 retains three-beat attainment with Sentence (8/16), but none with Words. LeVo2 yields at most one three-beat output per target-instruction condition. ACE-Step attains three beats in 4/16 cases with either its native field or Sentence alone, and none with Words alone. Other grouping labels remain in all denominators.

\begin{table}[htbp]
\centering\footnotesize
\setlength{\tabcolsep}{3pt}
\begin{tabular}{lllrrr}
\toprule Model & Form & Target & Acc & $\Delta_t$ & $R_t$\\\midrule
SA3 & Original & 3 & 0.438 [0.250,0.688] & 0.438 [0.250,0.688] & 0.063 [-0.250,0.375]\\
SA3 & Original & 4 & 0.438 [0.250,0.625] & -0.563 [-0.750,-0.375] & 0.250 [-0.063,0.500]\\
SA3 & Sentence & 3 & 0.500 [0.250,0.750] & 0.500 [0.250,0.750] & 0.313 [0.000,0.625]\\
SA3 & Sentence & 4 & 0.625 [0.438,0.750] & -0.375 [-0.563,-0.250] & 0.313 [0.000,0.563]\\
SA3 & Words & 3 & 0.000 [0.000,0.000] & 0.000 [0.000,0.000] & 0.000 [0.000,0.000]\\
SA3 & Words & 4 & 0.688 [0.500,0.875] & -0.313 [-0.500,-0.125] & -0.188 [-0.438,0.063]\\
\midrule
LeVo2 & Original & 3 & 0.000 [0.000,0.000] & 0.000 [0.000,0.000] & 0.000 [0.000,0.000]\\
LeVo2 & Original & 4 & 0.875 [0.688,1.000] & -0.125 [-0.313,0.000] & -0.063 [-0.188,0.000]\\
LeVo2 & Sentence & 3 & 0.063 [0.000,0.188] & 0.063 [0.000,0.188] & 0.063 [0.000,0.188]\\
LeVo2 & Sentence & 4 & 0.875 [0.750,1.000] & -0.125 [-0.250,0.000] & 0.063 [-0.125,0.250]\\
LeVo2 & Words & 3 & 0.000 [0.000,0.000] & 0.000 [0.000,0.000] & -0.063 [-0.188,0.000]\\
LeVo2 & Words & 4 & 0.938 [0.813,1.000] & -0.063 [-0.188,0.000] & 0.063 [0.000,0.188]\\
\midrule
ACE & Original & 3 & 0.250 [0.125,0.375] & 0.125 [0.000,0.313] & 0.188 [0.063,0.375]\\
ACE & Original & 4 & 0.875 [0.688,1.000] & 0.063 [0.000,0.188] & 0.375 [0.188,0.563]\\
ACE & Sentence & 3 & 0.250 [0.063,0.438] & 0.125 [-0.063,0.313] & 0.188 [0.063,0.375]\\
ACE & Sentence & 4 & 0.813 [0.625,0.938] & 0.000 [-0.188,0.188] & 0.125 [0.000,0.313]\\
ACE & Words & 3 & 0.000 [0.000,0.000] & -0.125 [-0.250,0.000] & 0.000 [0.000,0.000]\\
ACE & Words & 4 & 0.813 [0.625,0.938] & 0.000 [-0.125,0.125] & 0.063 [-0.125,0.250]\\\bottomrule
\end{tabular}
\caption{Independent one-seed expression experiment, 16 complete families per model. ACE Original uses the native time-signature field; Sentence/Words use text only. Sentence and Words denote the full time-signature sentence and ``Use three/four beats per bar.'' Exploratory pointwise 95\% intervals use 20,000 genre-stratified complete-family bootstrap draws; values are rounded to three decimals.}\label{tab:wording-two-model}
\end{table}

With Bonferroni correction across the 36 common model--form--target $\Delta_t/R_t$ comparisons, ACE native $R_4$ has interval [0.063,0.688]; all other common-condition $R_t$ intervals include zero. This indicates differing evidence strength, rather than proving an absence of discrimination in those conditions. Zero events or collapsed empirical intervals do not establish zero population probability.

\paragraph{Native-field-retained controls.}
ACE text-only Sentence/Words have four-beat $R_t$ values of 0.125/0.063; retaining the field yields 0.625/0.563 (Table~\ref{tab:wording-ace-field}). Three-beat attainment changes from 4/16 and 0/16 to 8/16 and 3/16. The eight additional $\Delta_t/R_t$ comparisons are corrected separately: Field+Sentence $R_3$ has interval [0.188,0.813]; Field+Sentence/Field+Words $R_4$ intervals are [0.250,0.938]/[0.250,0.875]. These results support a role for the native field in maintaining target selection; changes after removing it cannot all be attributed to wording.

\begin{table}[htbp]
\centering\footnotesize
\setlength{\tabcolsep}{3pt}
\begin{tabular}{lllrrr}
\toprule Model & Form & Target & Acc & $\Delta_t$ & $R_t$\\\midrule
ACE & Field+Sentence & 3 & 0.500 [0.250,0.750] & 0.375 [0.125,0.625] & 0.500 [0.250,0.750]\\
ACE & Field+Sentence & 4 & 0.938 [0.813,1.000] & 0.125 [0.000,0.250] & 0.625 [0.375,0.875]\\
ACE & Field+Words & 3 & 0.188 [0.063,0.375] & 0.063 [-0.125,0.250] & 0.188 [0.063,0.375]\\
ACE & Field+Words & 4 & 0.938 [0.813,1.000] & 0.125 [-0.063,0.313] & 0.563 [0.313,0.750]\\\bottomrule
\end{tabular}
\caption{ACE-Step paired controls retain the specified native time-signature field and append Sentence/Words to the caption. Each condition has 16 outputs. Table intervals are pointwise 95\%; the eight additional effect comparisons receive separate Bonferroni correction.}\label{tab:wording-ace-field}
\end{table}

\clearpage
\section{Target-Level Key Estimates}\label{app:key}
The following exploratory strata contain 16 target appearances per key, balanced across four genres. Intervals resample complete families within each genre; all target and model comparisons are not individually confirmatory tests. ``Other target'' varies according to the family pairing schedule. The table is generated from the same scoring-version file as Tables~\ref{tab:key}--\ref{tab:beat}.
\small
\begin{longtable}{llrrrr}
\toprule System & Key & Neutral & Acc & $\Delta$ & $R$\\\midrule
\endfirsthead
\toprule System & Key & Neutral & Acc & $\Delta$ & $R$\\\midrule
\endhead
ACE-Step & A major & 0.000 & 0.750 [0.563, 0.938] & 0.750 [0.563, 0.938] & 0.750 [0.563, 0.938] \\
ACE-Step & A minor & 0.188 & 0.813 [0.625, 0.938] & 0.625 [0.438, 0.813] & 0.813 [0.625, 0.938] \\
ACE-Step & Ab major & 0.000 & 0.750 [0.563, 0.938] & 0.750 [0.563, 0.938] & 0.438 [0.188, 0.688] \\
ACE-Step & Ab minor & 0.000 & 0.063 [0.000, 0.188] & 0.063 [0.000, 0.188] & 0.063 [0.000, 0.188] \\
ACE-Step & B major & 0.000 & 0.688 [0.438, 0.875] & 0.688 [0.438, 0.875] & 0.688 [0.438, 0.875] \\
ACE-Step & B minor & 0.063 & 0.938 [0.813, 1.000] & 0.875 [0.750, 1.000] & 0.813 [0.563, 1.000] \\
ACE-Step & Bb major & 0.000 & 0.438 [0.250, 0.625] & 0.438 [0.250, 0.625] & 0.438 [0.250, 0.625] \\
ACE-Step & Bb minor & 0.063 & 0.438 [0.250, 0.688] & 0.375 [0.188, 0.563] & 0.313 [0.063, 0.563] \\
ACE-Step & C major & 0.188 & 0.750 [0.563, 0.938] & 0.563 [0.313, 0.750] & 0.750 [0.563, 0.938] \\
ACE-Step & C minor & 0.000 & 0.813 [0.625, 1.000] & 0.813 [0.625, 1.000] & 0.813 [0.625, 1.000] \\
ACE-Step & C\# major & 0.000 & 0.500 [0.375, 0.625] & 0.500 [0.375, 0.625] & 0.500 [0.375, 0.625] \\
ACE-Step & C\# minor & 0.000 & 0.938 [0.813, 1.000] & 0.938 [0.813, 1.000] & 0.875 [0.625, 1.000] \\
ACE-Step & D major & 0.313 & 0.875 [0.688, 1.000] & 0.563 [0.375, 0.750] & 0.875 [0.688, 1.000] \\
ACE-Step & D minor & 0.000 & 0.938 [0.813, 1.000] & 0.938 [0.813, 1.000] & 0.938 [0.813, 1.000] \\
ACE-Step & E major & 0.125 & 0.563 [0.313, 0.813] & 0.438 [0.188, 0.688] & 0.563 [0.313, 0.813] \\
ACE-Step & E minor & 0.000 & 0.875 [0.688, 1.000] & 0.875 [0.688, 1.000] & 0.875 [0.688, 1.000] \\
ACE-Step & Eb major & 0.125 & 0.438 [0.250, 0.688] & 0.313 [0.125, 0.500] & 0.313 [0.063, 0.563] \\
ACE-Step & Eb minor & 0.000 & 0.563 [0.375, 0.688] & 0.563 [0.375, 0.688] & 0.563 [0.375, 0.688] \\
ACE-Step & F major & 0.063 & 0.125 [0.000, 0.313] & 0.063 [0.000, 0.188] & 0.063 [-0.125, 0.250] \\
ACE-Step & F minor & 0.000 & 0.438 [0.250, 0.688] & 0.438 [0.250, 0.688] & 0.438 [0.250, 0.688] \\
ACE-Step & F\# major & 0.000 & 0.813 [0.625, 1.000] & 0.813 [0.625, 1.000] & 0.813 [0.625, 1.000] \\
ACE-Step & F\# minor & 0.063 & 0.813 [0.625, 0.938] & 0.750 [0.563, 0.938] & 0.688 [0.438, 0.938] \\
ACE-Step & G major & 0.063 & 0.750 [0.563, 0.938] & 0.688 [0.438, 0.875] & 0.750 [0.563, 0.938] \\
ACE-Step & G minor & 0.063 & 0.938 [0.813, 1.000] & 0.875 [0.688, 1.000] & 0.938 [0.813, 1.000] \\
LeVo2 & A major & 0.063 & 0.000 [0.000, 0.000] & -0.063 [-0.188, 0.000] & 0.000 [0.000, 0.000] \\
LeVo2 & A minor & 0.063 & 0.063 [0.000, 0.188] & 0.000 [0.000, 0.000] & 0.063 [0.000, 0.188] \\
LeVo2 & Ab major & 0.000 & 0.000 [0.000, 0.000] & 0.000 [0.000, 0.000] & 0.000 [0.000, 0.000] \\
LeVo2 & Ab minor & 0.000 & 0.000 [0.000, 0.000] & 0.000 [0.000, 0.000] & 0.000 [0.000, 0.000] \\
LeVo2 & B major & 0.000 & 0.000 [0.000, 0.000] & 0.000 [0.000, 0.000] & -0.063 [-0.188, 0.000] \\
LeVo2 & B minor & 0.000 & 0.000 [0.000, 0.000] & 0.000 [0.000, 0.000] & 0.000 [0.000, 0.000] \\
LeVo2 & Bb major & 0.000 & 0.063 [0.000, 0.188] & 0.063 [0.000, 0.188] & -0.063 [-0.188, 0.125] \\
LeVo2 & Bb minor & 0.000 & 0.063 [0.000, 0.188] & 0.063 [0.000, 0.188] & 0.063 [0.000, 0.188] \\
LeVo2 & C major & 0.063 & 0.000 [0.000, 0.000] & -0.063 [-0.188, 0.000] & 0.000 [0.000, 0.000] \\
LeVo2 & C minor & 0.000 & 0.063 [0.000, 0.188] & 0.063 [0.000, 0.188] & 0.063 [0.000, 0.188] \\
LeVo2 & C\# major & 0.000 & 0.000 [0.000, 0.000] & 0.000 [0.000, 0.000] & 0.000 [0.000, 0.000] \\
LeVo2 & C\# minor & 0.000 & 0.000 [0.000, 0.000] & 0.000 [0.000, 0.000] & 0.000 [0.000, 0.000] \\
LeVo2 & D major & 0.000 & 0.000 [0.000, 0.000] & 0.000 [0.000, 0.000] & 0.000 [0.000, 0.000] \\
LeVo2 & D minor & 0.125 & 0.250 [0.125, 0.375] & 0.125 [0.000, 0.313] & 0.125 [0.000, 0.313] \\
LeVo2 & E major & 0.000 & 0.063 [0.000, 0.188] & 0.063 [0.000, 0.188] & 0.063 [0.000, 0.188] \\
LeVo2 & E minor & 0.063 & 0.063 [0.000, 0.188] & 0.000 [-0.188, 0.188] & 0.063 [0.000, 0.188] \\
LeVo2 & Eb major & 0.250 & 0.250 [0.063, 0.438] & 0.000 [-0.188, 0.250] & 0.250 [0.063, 0.438] \\
LeVo2 & Eb minor & 0.000 & 0.000 [0.000, 0.000] & 0.000 [0.000, 0.000] & 0.000 [0.000, 0.000] \\
LeVo2 & F major & 0.000 & 0.188 [0.000, 0.375] & 0.188 [0.000, 0.375] & 0.188 [0.000, 0.375] \\
LeVo2 & F minor & 0.000 & 0.000 [0.000, 0.000] & 0.000 [0.000, 0.000] & 0.000 [0.000, 0.000] \\
LeVo2 & F\# major & 0.000 & 0.063 [0.000, 0.188] & 0.063 [0.000, 0.188] & 0.000 [-0.125, 0.125] \\
LeVo2 & F\# minor & 0.000 & 0.000 [0.000, 0.000] & 0.000 [0.000, 0.000] & 0.000 [0.000, 0.000] \\
LeVo2 & G major & 0.000 & 0.125 [0.000, 0.313] & 0.125 [0.000, 0.313] & -0.063 [-0.188, 0.000] \\
LeVo2 & G minor & 0.063 & 0.063 [0.000, 0.188] & 0.000 [-0.125, 0.125] & -0.188 [-0.375, 0.000] \\
Stable Audio 3 & A major & 0.063 & 0.875 [0.688, 1.000] & 0.813 [0.625, 0.938] & 0.875 [0.688, 1.000] \\
Stable Audio 3 & A minor & 0.000 & 0.875 [0.688, 1.000] & 0.875 [0.688, 1.000] & 0.875 [0.688, 1.000] \\
Stable Audio 3 & Ab major & 0.000 & 0.188 [0.063, 0.375] & 0.188 [0.063, 0.375] & 0.188 [0.063, 0.375] \\
Stable Audio 3 & Ab minor & 0.000 & 0.125 [0.000, 0.250] & 0.125 [0.000, 0.250] & 0.125 [0.000, 0.250] \\
Stable Audio 3 & B major & 0.000 & 0.438 [0.188, 0.688] & 0.438 [0.188, 0.688] & 0.438 [0.188, 0.688] \\
Stable Audio 3 & B minor & 0.063 & 0.750 [0.563, 0.938] & 0.688 [0.500, 0.875] & 0.688 [0.438, 0.875] \\
Stable Audio 3 & Bb major & 0.000 & 0.750 [0.563, 0.938] & 0.750 [0.563, 0.938] & 0.688 [0.500, 0.875] \\
Stable Audio 3 & Bb minor & 0.000 & 0.188 [0.000, 0.375] & 0.188 [0.000, 0.375] & 0.188 [0.000, 0.375] \\
Stable Audio 3 & C major & 0.188 & 0.875 [0.750, 1.000] & 0.688 [0.500, 0.875] & 0.875 [0.750, 1.000] \\
Stable Audio 3 & C minor & 0.000 & 0.813 [0.625, 0.938] & 0.813 [0.625, 0.938] & 0.813 [0.625, 0.938] \\
Stable Audio 3 & C\# major & 0.000 & 0.438 [0.250, 0.625] & 0.438 [0.250, 0.625] & 0.438 [0.250, 0.625] \\
Stable Audio 3 & C\# minor & 0.000 & 0.938 [0.813, 1.000] & 0.938 [0.813, 1.000] & 0.688 [0.500, 0.875] \\
Stable Audio 3 & D major & 0.000 & 0.938 [0.813, 1.000] & 0.938 [0.813, 1.000] & 0.938 [0.813, 1.000] \\
Stable Audio 3 & D minor & 0.125 & 1.000 [1.000, 1.000] & 0.875 [0.688, 1.000] & 1.000 [1.000, 1.000] \\
Stable Audio 3 & E major & 0.000 & 0.875 [0.688, 1.000] & 0.875 [0.688, 1.000] & 0.875 [0.688, 1.000] \\
Stable Audio 3 & E minor & 0.000 & 0.875 [0.688, 1.000] & 0.875 [0.688, 1.000] & 0.875 [0.688, 1.000] \\
Stable Audio 3 & Eb major & 0.063 & 0.375 [0.188, 0.500] & 0.313 [0.125, 0.438] & 0.250 [0.000, 0.438] \\
Stable Audio 3 & Eb minor & 0.000 & 0.625 [0.438, 0.813] & 0.625 [0.438, 0.813] & 0.375 [0.188, 0.563] \\
Stable Audio 3 & F major & 0.000 & 0.438 [0.250, 0.688] & 0.438 [0.250, 0.688] & 0.438 [0.250, 0.688] \\
Stable Audio 3 & F minor & 0.000 & 0.875 [0.688, 1.000] & 0.875 [0.688, 1.000] & 0.875 [0.688, 1.000] \\
Stable Audio 3 & F\# major & 0.000 & 0.625 [0.500, 0.750] & 0.625 [0.500, 0.750] & 0.625 [0.500, 0.750] \\
Stable Audio 3 & F\# minor & 0.063 & 0.688 [0.500, 0.875] & 0.625 [0.438, 0.813] & 0.563 [0.250, 0.875] \\
Stable Audio 3 & G major & 0.125 & 0.813 [0.625, 0.938] & 0.688 [0.500, 0.875] & 0.813 [0.625, 0.938] \\
Stable Audio 3 & G minor & 0.000 & 0.813 [0.625, 1.000] & 0.813 [0.625, 1.000] & 0.750 [0.563, 0.938]
\\
\bottomrule
\end{longtable}

\bibliographystyle{plainnat}
\bibliography{references}
\end{document}